\documentclass{iau}
\usepackage{graphicx}

\title[Molecular gas and triggered star formation surrounding Wolf-Rayet stars]
{Molecular gas and triggered star formation surrounding Wolf-Rayet stars}

\author[Tie Liu, Yuefang Wu \& Huawei Zhang]   
{Tie Liu$^{1}$, Yuefang Wu$^{1}$ \& Huawei Zhang$^{1}$}

\affiliation{$^1$Department of Astronomy, Peking University, \\ Postbus 100871,
Beijing, China \\ email: {\tt liutiepku@gmail.com} }

\pubyear{2012}
\volume{292}  
\jname{Molecular Gas, Dust, and Star Formation in Galaxies}
\editors{Tony Wong \& Juergen Ott Editor, eds.}
\begin{document}

\maketitle

\begin{abstract}
The environments surrounding nine Wolf-Rayet stars were studied in molecular emission. Expanding shells were detected surrounding these WR stars (see left panels of Figure 1). The average masses and radii of the molecular cores surrounding these WR stars anti-correlate with the WR stellar wind velocities (middle panels of Figure 1), indicating the WR stars has great impact on their environments. The number density of Young Stellar Objects (YSOs) is enhanced in the molecular shells at $\sim$5 arcmin from the central WR star (lower-right panel of Figure 1). Through detailed studies of the molecular shells and YSOs, we find strong evidences of triggered star formation in the fragmented molecular shells (\cite[Liu et al.  2010]{liu_etal12}).

\end{abstract}

\firstsection 

\begin{figure}[thb]
\begin{center}
 \includegraphics[width=3in,angle=90]{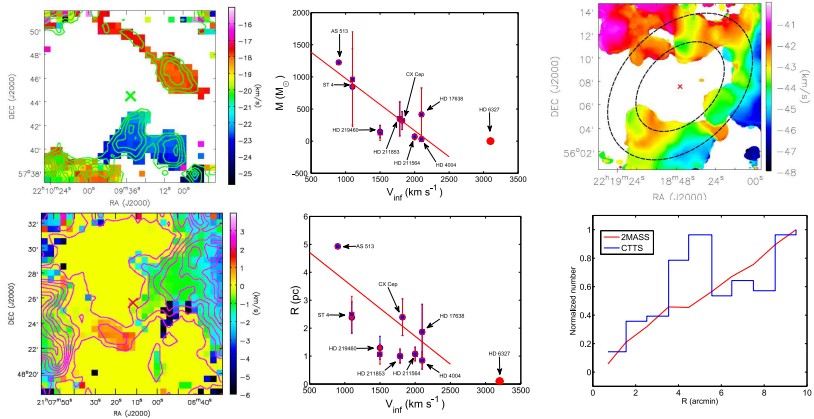}
 \caption{Upper left and Lower left: $^{12}$CO (1-0) integrated intensity contours on Moment 1 maps of WR stars CX Cep and ST 4, respectively. Middle panels: Average masses and radii of the molecular cores surrounding 9 WR stars as function of the stellar wind velocities. Upper right: $^{12}$CO (1-0) Moment 1 map of WR star HD 211853. Lower right: Radial bin number distribution of YSOs surrounding WR star HD 211853. (color version only available on line)  }
   \label{fig1}
\end{center}
\end{figure}

\end{document}